\documentclass[prl,twocolumn,showpacs,preprintnumbers,amsmath,amssymb,superscriptaddress]{revtex4-1}

\usepackage{graphicx}
\usepackage{dcolumn}
\usepackage{bm}
\usepackage{epsfig}
\usepackage{amsmath}
\usepackage{amssymb}
\usepackage{color}
\usepackage{bbm}
\usepackage[caption=false]{subfig}
\usepackage{placeins}
\usepackage{soul}

\usepackage[colorlinks=true,citecolor=blue,linkcolor=blue,urlcolor=blue]{hyperref}
\usepackage{cancel}

\def\equationautorefname~#1\null{%
	Eq.~#1\null
}
\def\figureautorefname~#1\null{%
	Fig.~#1\null
}

\begin{document}

\title{Out-of-Time-Order Correlation as a Witness for Topological Phase Transitions}
\author{Qian Bin}
\affiliation{School of Physics and Institute for Quantum Science and Engineering, Huazhong University of Science and Technology, Wuhan, 430074, China}

\author{Liang-Liang Wan}
\affiliation{School of Physics and Institute for Quantum Science and Engineering, Huazhong University of Science and Technology, Wuhan, 430074, China}

\author{Franco  Nori}
\affiliation{Theoretical Quantum Physics Laboratory, RIKEN Cluster for Pioneering Research, Wako-shi, Saitama 351-0198, Japan}
\affiliation{RIKEN Center for Quantum Computing (RQC), 2-1 Hirosawa, Wako-shi, Saitama 351-0198, Japan }
\affiliation{Physics Department, The University of Michigan, Ann Arbor, Michigan 48109-1040, USA}

\author{Ying Wu}
\affiliation{School of Physics and Institute for Quantum Science and Engineering, Huazhong University of Science and Technology, Wuhan, 430074, China}

\author{Xin-You L\"{u}}
\email{xinyoulu@hust.edu.cn}
\affiliation{School of Physics and Institute for Quantum Science and Engineering, Huazhong University of Science and Technology, Wuhan, 430074, China}

\date{\today}

\begin{abstract}
We propose a physical witness for dynamically detecting topological phase transitions (TPTs) via an experimentally observable out-of-time-order correlation (OTOC). The distinguishable OTOC dynamics appears in the topological trivial and non-trivial phases due to the topological locality. In the long-time limit, the OTOC undergoes a {\it zero-to-finite-value transition} at the critical point of the TPTs. This transition is robust to the choices of the initial state of the system and the used operators in OTOC. The proposed OTOC witness can be applied into the systems with and without chiral symmetry, e.g., the lattices described by the SSH model, Creutz model, and Haldane model. Moreover, our proposal, as a physical witness in real space, is still valid even in the presence of disorder. Our work fundamentally offers a new prospect of exploring topological physics with quantum correlations.
\end{abstract}
 
\maketitle
Topological phase transitions (TPTs) are fundamentally interesting in modern physics because these go beyond the paradigm of traditional phase transitions associated with symmetry breaking\,\cite{Bansil2016LD}.  It offers a non-trivial paradigm for the classification of matter phases, and thus is attracting enormous attention in condensed matter physics\,\cite{Hasan2010Kane,Qi2011Zhang,Bernevig2013Hughes,Chiu2016TS}, optics\,\cite{Ozawa2019PA}, and non-Hermition physics\,\cite{Bergholtz2021BK}. The occurrence of TPTs involve the gap-closing-and-opening of band (the change of system topology) with symmetry preserving. According to the extended bulk-boundary correspondence, the $n$th-order TPT in a $d$-dimensional (dD) system leads to the appearance of a $(d-n)$-dimensional gapless boundary state in the topological non-trivial phase\,\cite{Langbehn2017PT,Schindler2018CV,Liu2019ZA,Kunst2018MB,Imhof2018BB,Ezawa2018,Schindler2018WV,Noh2018BH,Geier2018TH,Chen2018CG, Zhu2021HZ,Che2020GL}. This symmetry-protected boundary state has strong robustness to disorder\,\cite{Mittal2014FF, MondragonShem2014HS,Meier2018AD} and defects\,\cite{Bandres2018WH}. It can be used to realize topological lasers exhibiting robust transports\,\cite{StJean2017GG,Zhao2018MT,Parto2018WH,Harari2018BL, Bandres2018WH}, topological protected quantum coherence\,\cite{Bahri2015VA, Nie2020PN}, and quantum state transfer\,\cite{Yao2013LG}. Thus, the detection of TPTs is a key for exploring topological physics. To quantitatively distinguish the topological trivial and non-trivial phases, normally one calculates topological invariants (e.g., winding number and Chern number) in momentum space\,\cite{Asboth2016OP}. However, identifying TPTs with those commonly used topological invariants is not suited for disorder systems  where it is difficult to give the Hamiltonian in momentum space. Then, it becomes a significant task to identify TPTs via an alternative physical witness in real space that is robust to disorder.

The OTOC, defined as $\mathcal{O}(t)=\langle W^\dag(t)V^\dag W(t) V\rangle$ with $W(t)=e^{iHt}We^{-iHt}$, was proposed in investigating the holographic duality between a strongly interacting quantum system and a gravitational system\,\cite{Shenker2014Stanford,Shenker2015Stanford,Maldacena2016SS,Jensen2016,Roberts2016Swingle,Garcia-Garcia2018LRB}.  Here $W$ and $V$ are initially commuting operators\,\cite{Swingle2018}.  Different from the normal time-order correlation function characterizing classical and  quantum statistics\,\cite{Glauber1963,Scully1997,Agarwal2013,Salmon2022Gustin,Mercurio2022Macri,Lin2021LK}, the OTOC can quantify the temporal and spatial correlations throughout many-body quantum systems,  which is closely related to information scrambling. Thus, it is a widely used tool for diagnosing chaotic behavior\,\cite{Roberts2015Stanford,Gu2016Qi,Zhu2016HG,Garcia-Mata2018SJ,Aleinera2016FI,Rozenbaum2017GG,Nahum2018VH,Khemani2018VH,Keyserlingk2018RP,Chan2018LC,Rozenbaum2019GG,Cao2020ZS,Liao2018Galitski,Syzranov2018GG,Lin2018Motrunich,Akutagawa2020HS,Xu2020SC,Alonso2022SA,Blocher2020AM},  many-body localization\,\cite{Huang2017ZC,Chen2017ZH,Fan2017ZS,He2017Lu,Gu2017,Lee2019KK,Swingle2017Chowdhury,Smit2019KM}, entanglement\,\cite{McGinley2019NK,Riddell2019Sorensen,Luitz2017Lev,Garttner2018HR,Hosur2016QR}, and quantum phase transitions\,\cite{Shen2017ZF,Heyl2018PD,Dag2019SD,Chen2020HZ,Sun2020CT,Wang2019PerezBernal,LewisSwan2020MR,Dag2020DS}. Here, many-body localization is a kind of many-body phenomenon in the nonequilibrium system caused by many-body interactions. This is essentially different from TPTs that describe the change of topological structure of systems. Under the frame of band topology theory, normally the TPTs occurs in the system without the many-body interactions.   Recently, the dynamical detection of TPTs with OTOCs has been proposed\,\cite{Dag2020DS}, where the infinite-temperature OTOCs  can directly probe zero-temperature quantum phases via detecting the presence of Majorana zero modes.  This construct a relation between infinite-temperature information scrambling and zero-temperature $\mathbb{Z}_2$ topological order. 
Moreover, the OTOC can also be implemented experimentally\,\cite{Garttner2017BSN,Li2017FW,Landsman2019FS,Nie2020WC,Wei2019PS} by connecting the time reversal to the Loschmidt echo technique\,\cite{Hahn1950SE,Swingle2016BSS,Sanchez2020SC}. Further exploiting OTOC dynamics in topological systems may open a door for completing the challenging problem of identifying TPTs in the presence of disorder. Until now, the relation between zero-temperature OTOC and TPTs remains largely unexplored, which may substantially advance the fields of quantum correlation and topological physics.

Here we propose a zero-temperature OTOC witness for dynamical detecting $\mathbb{Z}$-type TPTs in lattice systems. As shown in Fig.\,\ref{fig1}(a), the constructed OTOC becomes an experimentally observable fidelity~\cite{Garttner2017BSN} of a final state $\rho_f$ projected onto an initial state $\rho_0$ by defining $V=V\rho_0=|\psi_0\rangle\langle\psi_0|$, i.e., 
\begin{align}\label{eq01}
\mathcal{O}(t)={\rm tr}[\rho_0e^{iHt}W^\dag e^{-iHt}\rho_0 e^{iHt}We^{-iHt}]=F(t).
\end{align}
Due to the topological locality, the long-time limit of the OTOC $\mathcal{O}(t\rightarrow\infty)$ undergoes a {\it zero-to-finite-value transition} along with the system entering into the non-trivial phase from the trivial phase. This sudden change is not limited by the choices of the operators $V$ (corresponding to the initial state of system) and $W$. In comparison with previous methods of detecting TPTs\,\cite{Chiu2016TS}, the proposed OTOC, as a witness in real space, can be applied in {\it disordered systems}. Moreover, it is not only suitable for the systems with chiral symmetry described by the nearest-neighbor (NN) Su-Schrieffer-Heeger (SSH) model, next-next-nearest-neighbor (NNNN) SSH model and Creutz model, but also can be used to the systems without chiral symmetry, such as 2D lattices described by the Haldane model and Qi-Wu-Zhang model. We also demonstrate the validity of the OTOC witness for detecting second-order TPTs. 
\begin{figure}
\includegraphics[width=8.6cm]{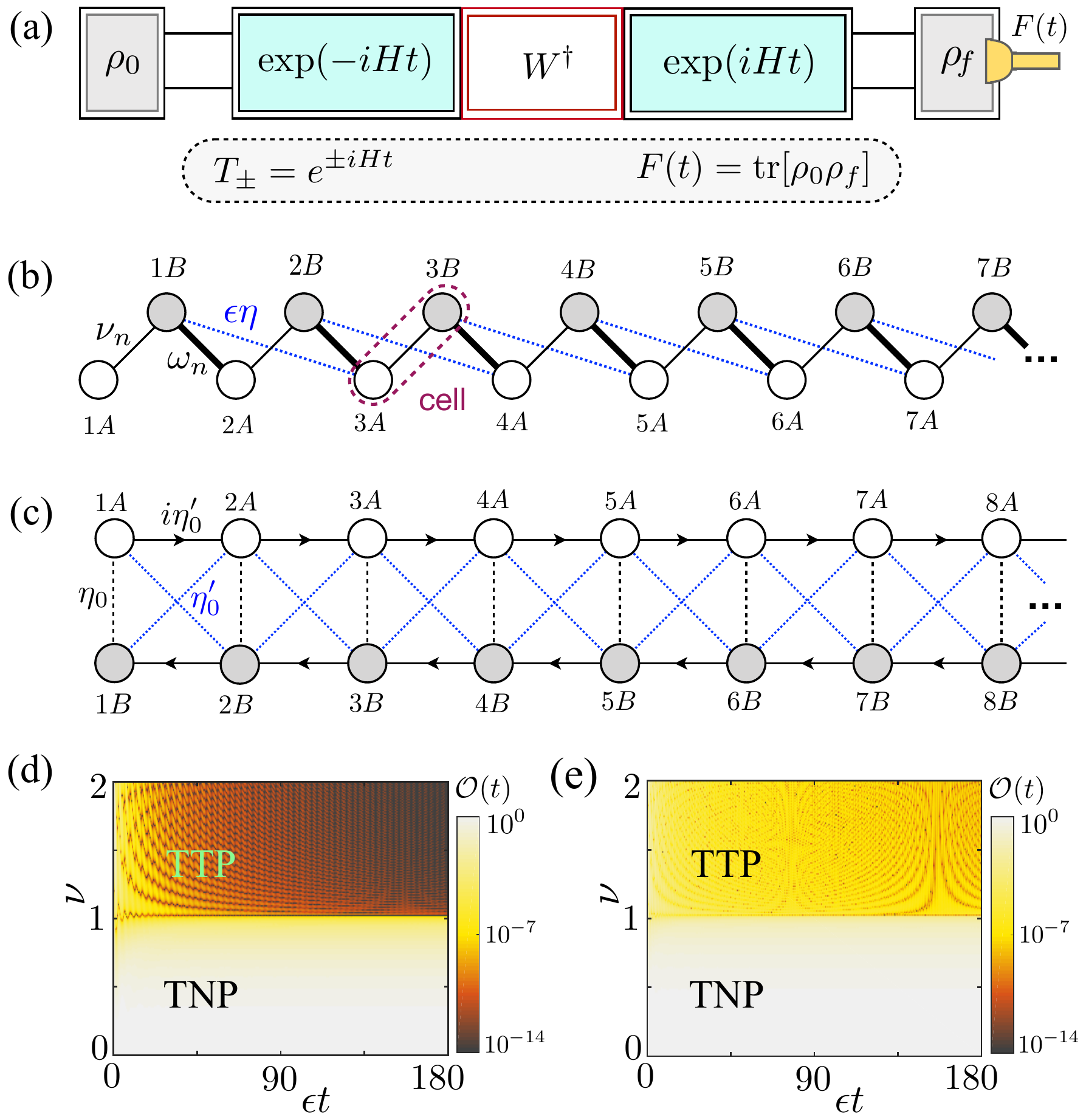}\\
\caption{(a) A schematic illustration of implementing the OTOC, which is equal to the fidelity $F(t)={\rm tr}[\rho_0\rho_f]$\,\cite{Garttner2018HR, Garttner2017BSN}. First, the initial state $\rho_0$ evolves to the state $\rho_1(t)$ under $T_{-}=e^{-iHt}$. Second, the system changes from $\rho_1(t)$ to $\rho_2(t)$ after the operation of $W$. Lastly, the system evolves backward  to get the final state $\rho_f$ under $T_{+}=e^{iHt}$. (b, c) Schemes of the 1D SSH model and Creutz model, which describe the lattice systems with chiral symmetry. (d, e) Phase diagrams of the NN SSH model: the OTOC versus $\epsilon t$ and $\nu$ for (d) $W=a_{1,A}^\dag a_{1,A}$ and (e) $W=\sum_{n=1}^{N-1} a_n^\dag\sigma_3 a_n$, where $N=200$, $|\psi_0\rangle=|1, A\rangle$, and $d_1=d_2=0$. The topological non-trivial and trivial phases are denoted as TNP and TTP, respectively.}\label{fig1}
\end{figure}

\emph{Detecting TPTs in the systems with chiral symmetry.}---Without loss of generality, we choose the 1D SSH model and Creutz model depicted in Figs.\,\ref{fig1}(b,c) as examples for demonstrating the validity of detecting TPTs with OTOC in the systems with chiral symmetry. The corresponding system Hamiltonians can be written as\,\cite{Asboth2016OP,Su1979SH,Rufo2019LC,Creutz1999}
\begin{subequations}
\begin{align}\label{eq02}
&\!\!H_{\rm s} \!=\!\!\! \sum_{n}\!\{ \nu_n a_n^\dag \sigma_1 a_n\!\!+\!\![ (\omega_n a_{n+1}^\dag\!\!+\!\epsilon \eta a_{n+2}^\dag)\frac{\sigma_1\!\!+\!\!i\sigma_2}{2}a_{n}\! \!+\! {\rm h.c.}   ]   \},\\
&\!\!H_{\rm cr} \!=\! \!\sum_{n} \{ \eta_0 a_n^\dag \sigma_1 a_n +\eta_0'[ a_{n+1}^\dag \frac{\sigma_1\!-\!i\sigma_3}{2}a_n+{\rm h.c.}] \},
 \end{align}
 \end{subequations}
where the number of cells is $N$, $\sigma_j$ ($j=0, 1, 2, 3$) is Pauli operator, and $a_n^\dag=(a_{n,A}^\dag, a_{n,B}^\dag)$ is the annihilation operator of the unit cell $n$ with sublattices $A$, $B$. For the SSH model with Hamiltonian $H_{\rm s}$, $\omega_n=\epsilon(1+d _1r_n)$ [or $\nu_n=\epsilon(\nu+d_2 r_n')$] is the intercell (or intracell) hopping strength. Disorder with the dimensionless strengths $d_1$, $d_2$ has been included here, and $r_n$, $r_n'$ are the independent random real numbers chosen from the uniform distribution $[-0.5,0.5]$. Physically, $\epsilon$ is the characteristic intercell strength, $\nu$ is the ratio of intra- to inter-cell hopping in the clean system, and $\epsilon\eta$ is the NNNN hopping strength. Here, $H_{\rm s}$ is reduced to a standard Hamiltonian of the NN SSH model when $\eta=0$. For the Creutz model with Hamiltonian $H_{\rm cr}$, the arrows indicate the sign of the hopping phase, and $\eta_0$ ($\eta_0'$) is the vertical (horizontal and diagonal) hopping strength. The above models possess a chiral symmetry with a well-defined chiral operator $\mathcal{C}_{\rm 1d}$, which can reverse the energy of the system, i.e., $\mathcal{C}_{\rm 1d}H\mathcal{C}_{\rm 1d}^{-1}=-H$ ($H=H_{\rm s}, H_{\rm cr}$), where $\mathcal{C}_{\rm 1d}=\sum_{n=1}^N a_n^\dag\sigma_3 a_n$ for the SSH model and $\mathcal{C}_{\rm 1d}=\sum_{n=1}^N a_n^\dag\sigma_2 a_n$ for the Creutz model.

Let's first consider the case of no disorder, i.e., $d_1=d_2=0$, the NN (and NNNN) SSH model and Creutz model feature the TPTs at $\nu=1$ (and $\eta=0,1$) and $\eta_0=\eta'_0$, respectively\,\cite{Asboth2016OP,Su1979SH,Rufo2019LC,Creutz1999}. To identify the topological non-trivial and trivial phases in real space, in Fig.\,\ref{fig2}, we numerically calculate the OTOC dynamics with Eq.\,(\ref{eq01}), which involves the backward evolution. Note that, Fig.\,\ref{fig2} includes the results for choosing different OTOC operators $V$ and $W$. It clearly shows that, both for the SSH model and Creutz model, the distinguishable OTOC dynamics appears in the non-trivial and trivial phases. Specifically, the OTOC evolves to a finite value and almost zero in the topological non-trivial and trivial phases, respectively [see the insets of Figs.\,\ref{fig2}(b,d,f)]. This relates to the physical mechanism that the information does scramble in the trivial phase, while this scrambling is suppressed immensely in the nontrivial phase. There exists a {\it zero-to-finite-value transition} in the long-time limit of the OTOC, when the system enters into the non-trivial phase from the trivial phase. This distinguishable OTOC dynamics is robust to the initial state of the system (i.e., the operator $V$), which could be a single-site occupation or  multi-site occupation state. Moreover, the averaged OTOC becomes discrete at the critical point, when the initial state is the eigenstate of the system whose eigenvalue has the lowest absolute value~\cite{supp}. Figure\,\ref{fig2} also shows that the OTOC witness is not limited by the choice of the operator $W$. In our proposal, the operator $W$ can either a few-site (including single-site) operation on sublattice $A$ (e.g., $W=\sum_{l=1}^{L}a_{l,A}^\dag a_{l,A}$, $L=1,2,3$) or a multi-site operation on sublattices $A$ and $B$ (e.g., $W=\sum_{n=1}^{N-1} a_n^\dag\sigma_j a_n$, $j=2,3$), and the chosen operators $W$ neither commute nor anti-commute with the system Hamiltonian, i.e., $[W,H]_{\pm}\neq0$. 
\begin{figure}
\includegraphics[width=8.5cm]{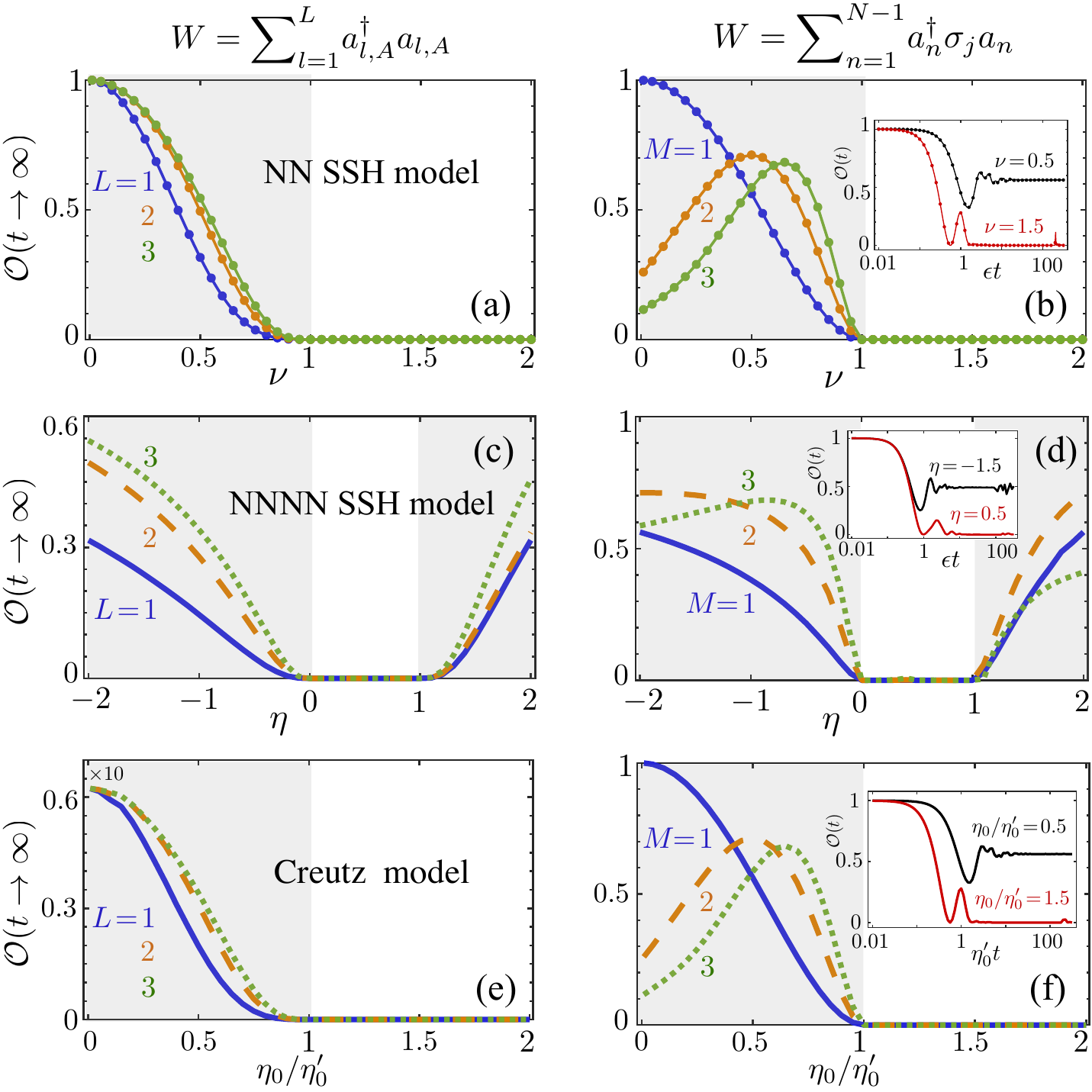}\\
\caption{The dependence of $\mathcal{O}(t\rightarrow\infty)$ on $\nu$, $\eta$, and $\eta_0/\eta_0'$ for (a,c,e) $W=\sum_{l=1}^L a_{l,A}^\dag a_{l,A}$ and (b,d,f) $W=\sum_{n=1}^{N-1} a_n^\dag\sigma_j a_n$ [$j=3$ for (b,d) and $j=2$ for (f)]. Panels (a,b), (c,d), and (e,f) correspond to the systems described by the NN SSH model, NNNN SSH model, and Creutz model, respectively. The initial states are set as (a,c,e) $|\psi_0\rangle= |1,A\rangle$, (b,d) $|\psi_0\rangle=\sum_{m=1}^M (-1)^{m-1} |m,A\rangle/\sqrt{M}$ and (f) $|\psi_0\rangle=\sum_{m=1}^M (-1)^{m-1}( |m,A\rangle+i|m,B\rangle)/\sqrt{2M}$. Insets: the evolution of the OTOC for different values of $\nu$, $\eta$ and $\eta_0/\eta_0'$ when $M=1$. The lines and dots correspond to the fully numerical simulations obtained by Eq.\,(\ref{eq01}) and the analytical results obtained by Eqs.\,(\ref{eq03},\ref{eq04}), respectively.  Other system parameters are $N=200$, $d_1=d_2=0$, (a,b) $\eta=0$, (c,d) $\nu=1$. The TNPs and TTPs are indicated by the gray shadings and write areas, respectively.}\label{fig2}
\end{figure}

To fully show the dependence of the OTOC witness on system parameters, we also calculate the analytical solution of $\mathcal{O}(t)$ under the condition of $N\gg1$. Let's consider the NN SSH model as an example, and choose $|\psi_0\rangle=\sum_{m=1}^M \frac{(-1)^{m-1}}{\sqrt{M}}|m,A\rangle$, where $M=1$ corresponds to the case of single-site occupation state, i.e., $|\psi_0\rangle=|1,A\rangle$. Here, $m$ and $A/B$ in state $|m,A/B\rangle$ represent the $m$th cell and sublattice $A/B$, respectively. Corresponding to $W= \sum_{l=1}^L a_{l,A}^\dag a_{l,A}$ and $W=\sum_{n=1}^{N-1} a_n^\dag\sigma_3 a_n$, we respectively obtain\,\cite{supp}
\begin{align}\label{eq03}
\!\!\mathcal{O}(t)\!\approx\![1/\!\sum_{n=0}^N\nu^{2n}\!+\!\sum_{k=1}^N\!\frac{2 \epsilon^2\nu^2 \cos(\lambda_+^{(k)}t) }{(N+1)(\lambda_{\pm}^{(k)})^2}\sin^2(\!\frac{k\pi}{N+1}\!)]^4
\end{align}
and
\begin{align}\label{eq04}
\!\!\mathcal{O}(t)\!\approx\! [1/\!\sum_{n=0}^N\!\nu^{2n}\!\!+\!\!\sum_{k=1}^N\!\frac{2 \epsilon^2\nu^2 \cos(2\lambda_+^{(k)}t) }{(N+1)(\lambda_{\pm}^{(k)})^2}\sin^2(\!\frac{k\pi}{N+1}\!)]^2
\end{align}
for $L,M=1$. Here $\lambda_\pm^{(k)}=\pm \epsilon[1+\nu^2+2\nu  \cos(\frac{k\pi}{N+1})]^{1/2}$ and $k=1,2,\dots, N$.  Note that the above equations require $\nu\neq 0$, and $\nu=0$ means that the hopping cannot occur in the intracells, corresponding to $\mathcal{O}(t)=1$. The similar analytical results for $L,M>1$ are shown in the supplementary material\,\cite{supp}.  As shown in Figs.\,\ref{fig1}(a,b), the analytical solutions also present a {\it zero-to-finite-value transition} of OTOC at the critical point of TPTs. This conclusion is valid for both the cases of choosing $W$ as a single-site operation and a multi-site operation. Figures \ref{fig2}(a,b) show a very good agreement between the analytical solutions  and the fully numerical simulations, which demonstrates the validity of our solutions.      

Now let's discuss the influence of disorder on our proposal by choosing the NN SSH model as an example. The proposed OTOC witness for identifying the TPTs is also suitable for {\it disordered systems}. As shown in Figs.\,\ref{fig3}(a,b), $\mathcal{O}(t\rightarrow\infty)$ still undergoes the {\it zero-to-finite-value transition} along with the occurrence of the TPTs, even when weak disorder is introduced into the system.  In terms of information, this transition originally comes from the topological locality in the non-trivial phase. Specifically, the information scrambling occurs in the trivial phase, and is suppressed immensely in the non-trivial phase. Similar as the case of no disorder, this result is robust to the choices of the operator $W$. Figures\,\ref{fig3}(a,b) also show that the above distinguishability of the OTOC dynamics disappears in the strong disorder regime (e.g., $d>4$). Physically, this is because the TPTs, together with the symmetry-protected boundary state, will disappear as the disorder is too large. Figures \ref{fig3}(c,d) further demonstrate the vanishing of the topological non-trivial phase induced by strong disorder. Moreover, the proposed OTOC witness can also be considered as an order parameter of the topological phase diagram, and predict topological Anderson insulator physics~\cite{supp}. It is consistent with previous works in Refs.\,\cite{MondragonShem2014HS,Meier2018AD}, which further verify the validity of our OTOC witness.  
 \begin{figure}
\includegraphics[width=8.2cm]{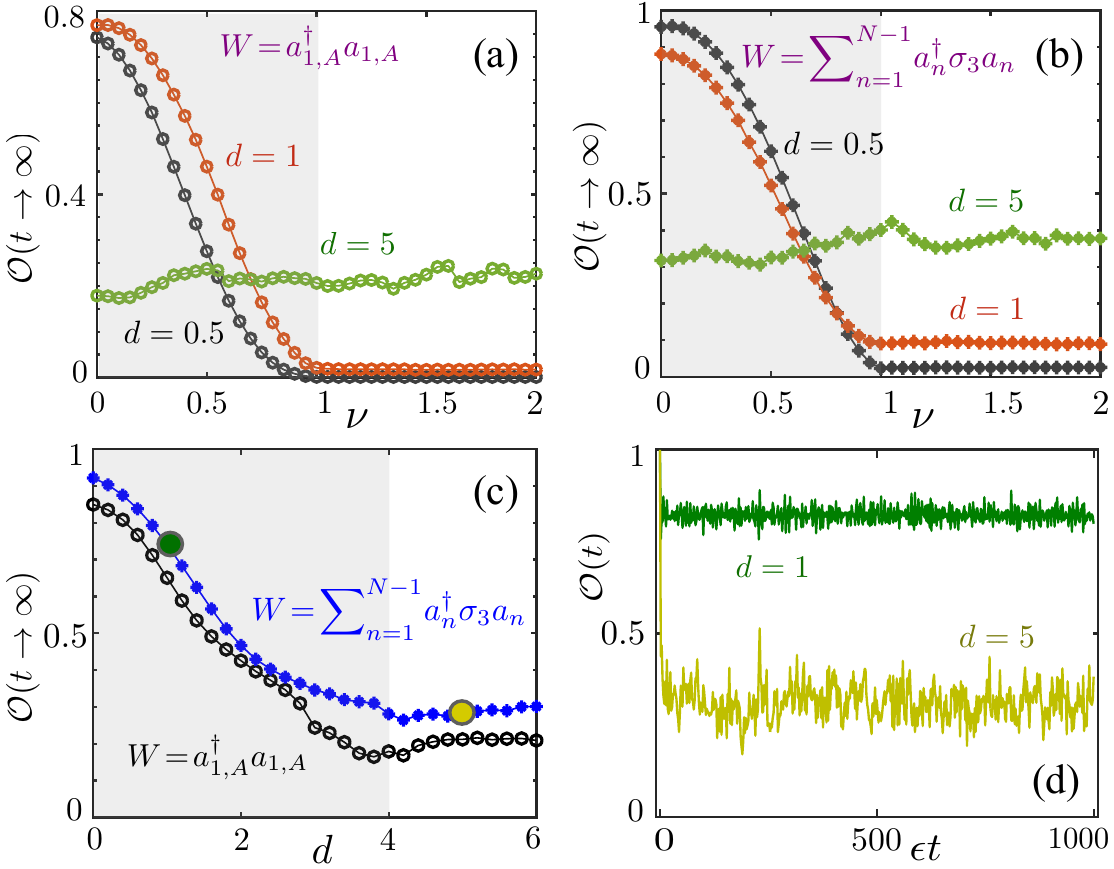}\\
\caption{ (a,b) The dependence of $\mathcal{O}(t\rightarrow\infty)$ on $\nu$ for different  disorder strengths $d$ when (a) $W=a_{1,A}^\dag a_{1,A}$ and (b) $W=\sum_{n=1}^{N-1} a_n^\dag\sigma_3 a_n$. (c) The value of $\mathcal{O}(t\rightarrow\infty)$ versus $d$ for different choices of the operator $W$ when $\nu=0.2$. (d) The evolution of the OTOC for different $d$ indicated by the circles in (c). Here all data are averaged over 30 independent disorder configurations, and we have chosen $N=200$, $d_2=2d_1=d$, and $|\psi_0\rangle=|1,A\rangle$. The TNPs and TTPs are indicated by the gray shadings and write areas, respectively.}\label{fig3}
\end{figure}

\emph{Detecting TPTs in the systems without chiral symmetry.}---The proposed OTOC witness for identifying the TPT is not limited to the above systems with chiral symmetry, but is applicable for the systems without chiral symmetry, such as 2D lattice systems described by the Haldane model and Qi-Wu-Zhang model. As shown in Fig.\,\ref{fig4}(a), the Haldane model on the honeycomb lattice has Hamiltonian\,\cite{Haldane1988,Jotzu2014MD} 
\begin{align}\label{eq05}
H_{\rm ha}\!=\!\eta_1\!\sum_{\langle j,j'\rangle}c_{j}^\dag c_{j'}\!+\!\eta_2\!\!\sum_{\langle\langle j,j'\rangle\rangle}\!e^{is_{jj'}\phi}c_{j}^\dag c_{j'} \!+\!\mu s' \!\sum_j c_{j}^\dag c_{j},
 \end{align}
where $c_j^\dag$ ($c_j$) is the creation (annihilation) operator of the $j$th site, and the summation indexes cover all sites. The symbol $\mu$ in last term denotes the sublattice potential, where $s'=+1$ and $s'=-1$ correspond to sublattices $A$ and $B$, respectively. Here, $\eta_1$ and $\eta_2$ are the real-valued nearest- and next-nearest-neighbor hopping amplitudes, respectively.  The next-nearest-neighbor hopping contains the phases $s_{jj'}\phi$ with $s_{jj'}=\pm1$, which  can break the time-reversal symmetry. The system has no chiral symmetry and is a paradigmatic example of 2D lattice featuring TPTs. For example, the parameter ranges $|\mu/\eta_2|<3\sqrt{3}$ and $\mu/\eta_2=other$ correspond respectively to the topological non-trivial and trivial phases when $\phi=\pi/2$. Similar as the procedure used in 1D systems with chiral symmetry, we numerically calculate the OTOC dynamics with Eq.\,(\ref{eq01}) to identify the occurrence of TPTs in real space. As shown in Fig.\,\ref{fig4}(b), the {\it zero-to-finite-value transition} of $\mathcal{O}(t\rightarrow\infty)$ can still be observed when the system enters into the topological nontrivial phase from the trivial phase. The similar results can also be obtained in the system described by the Qi-Wu-Zhang model\,\cite{supp}.

\begin{figure}
\includegraphics[width=8.6cm]{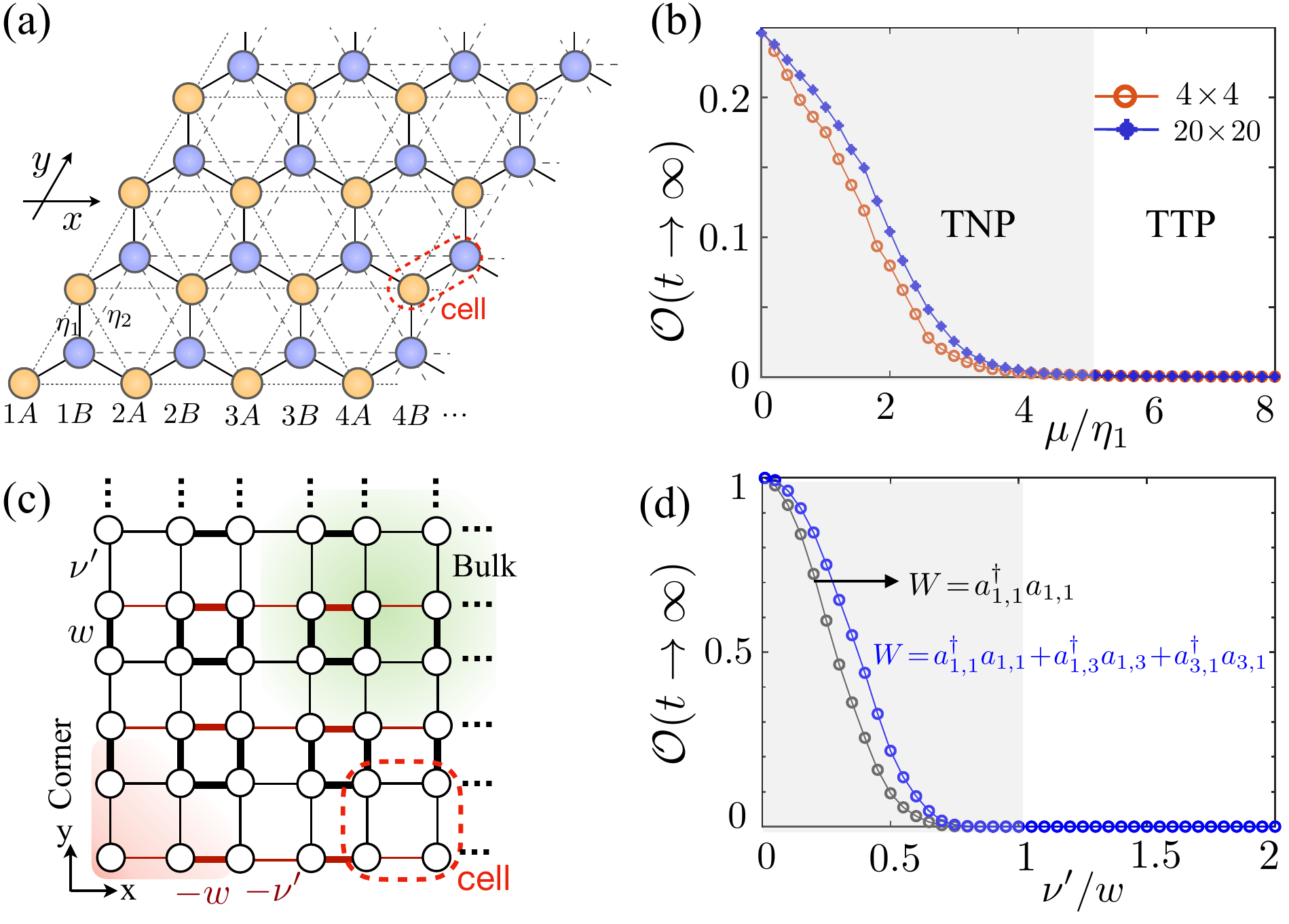}\\
\caption{(a) Scheme of the Haldane model, where the unit cell consists of sublattices $A$ and $B$. (b) The dependence of $\mathcal{O}(t\rightarrow\infty)$ on $\mu/\eta_1$ for different cell numbers when  $|\psi_0\rangle=|1,A\rangle$ and $W=\sum_{j} c_{j}^\dag c_{j}$ (the summation index $j$ only cover all sublattice $B$). Here we have chosen $\eta_1=\eta_2$ and $\phi=\pi/2$. The red and blue lines correspond to the cell numbers of $4\times 4$ and $20\times 20$, respectively.  (c) Scheme of the 2D SSH model with gauge flux $\pi$ penetrating any plaquette. (d) The dependence of $\mathcal{O}(t\rightarrow\infty)$ on $\nu'/w$ for $W=a_{1,1}^\dag a_{1,1}$ (black curve) and $W=a_{1,1}^\dag a_{1,1}+a_{1,3}^\dag a_{1,3}+a_{3,1}^\dag a_{3,1}$ (blue curve) when  $|\psi_{0}\rangle=|1, 1\rangle$. The TNPs and TTPs are indicated by the gray shadings and write areas, respectively.}\label{fig4} 
\end{figure}

\emph{Application to the second-order TPTs}.---Higher-order topological insulators, as an extension of the  topological insulators, have recently attracted extensive attention\,\cite{Langbehn2017PT,Schindler2018CV,Liu2019ZA,Kunst2018MB,Imhof2018BB,Ezawa2018,Schindler2018WV,Noh2018BH,Geier2018TH,Chen2018CG,Zhu2021HZ,Che2020GL}. High-order TPTs usually can be identified by detecting the boundary states in real space. For example, the topological protected corner states have been used to identify the second-order TPT in a 2D system\,\cite{Benalcazar2017BH,SerraGarcia2018PS,Peterson2018BH,Mittal2019OZ}. Here, our proposed OTOC witness is also applicable for detecting second-order TPTs. As shown in Fig.\,\ref{fig4}(c), we take the extended 2D SSH model with non-zero gauge flux as an example, and its Hamiltonian reads\,\cite{Mittal2019OZ}
\begin{align}\label{eq6}
\!\!H_{\rm 2s}({\rm {\bf k}})=&(\nu'+w\cos k_y)\tau_0 \otimes \sigma_1 \!-\!w\sin k_y\tau_3\otimes\sigma_2\nonumber\\
&-(\nu'+w\cos k_x)\tau_2\otimes\sigma_2\!-\!w\sin k_x\tau_1\otimes\sigma_2,
\end{align}
where ${\rm {\bf k}}=\{k_x, k_y\}$ are the wave number, and $\pm\nu'$ ($\pm w$) is the intracell (intercell) hopping strength. This system features a second-order TPT when increasing the value of $\nu'/w$, i.e., $\nu'\!<\!w$ and $\nu'\!>\!w$ corresponding to the topological non-trivial and trivial phases, respectively. To identify the occurrence of second-order TPTs, in Fig.\,\ref{fig4}(d), we numerically calculate the OTOC in the lattice system with $20\times20$ cells when the different OTOC operators $W$ are considered. Figure \ref{fig4}(d) clearly shows the distinguishable OTOC dynamics in the topological non-trivial and trivial phases. Both for $W=a_{1,1}^\dag a_{1,1}$ and $W=a_{1,1}^\dag a_{1,1}+a_{1,3}^\dag a_{1,3}+a_{3,1}^\dag a_{3,1}$, the {\it zero-to-finite-value transition} of $\mathcal{O}(t\rightarrow\infty)$ appears at the critical point of the second-order TPT. Moreover, the system is initially in the corner site $(1,1)$ (i.e., $|\psi_0\rangle=|1,1\rangle$), which is experimentally feasible. Here $(x, y)$ represents a lattice point in the square lattice, and $|x, y\rangle$ denotes the state occupying in the site $(x,y)$. The creation (annihilation) operator of the  site $(x,y)$ is denoted by $a_{x,y}^\dag$ ($a_{x,y}$).        

\emph{Experimental implementation and conclusions.}---Regarding experimental implementations, the trapped ion\,\cite{Nevado2017FLP,Garttner2017BSN,Haffner2008RB,Blatt2012Roos,Monroe2021CD} is an ideal candidate for our proposal. We consider a set of $2N$ trapped ions with excited and ground states arranged along a 1D chain as the SSH model. First, the system is initialized in $\rho_0=|1,A\rangle\langle1,A|$ by applying a $\pi$ pulse to excite the first ion in the chain into its excited state\,\cite{Haffner2008RB,Blatt2012Roos,Monroe2021CD}. Then, one should make the system evolve under the Hamiltonian for a time $t$ to the state $\rho_1(t)=e^{-iHt}\rho_0e^{iHt}$. Subsequently, applying the operator $W$ to get $\rho_2(t)=W^{\dagger}\rho_1(t)W $. When the operator $W$ is a single-site operator on sublattice $A$, it can be achieved by removing the polarizations of the ions except for that of the first ions by using selective pulses\,\cite{Haffner2008RB,Blatt2012Roos,Monroe2021CD,Garttner2017BSN}. Next,  inverting the sign of $H$ by the spin echo technique (i.e., applying a $\pi$ pulse to reverse the polarization of one of the ions)\,\cite{Hahn1950SE} and making the system evolve again for $t$ to obtain the final state $\rho_f=e^{iHt}\rho_2(t)e^{-iHt}$\,\cite{Swingle2016BSS,Sanchez2020SC}. Finally, the OTOC can be obtained by measuring the overlap of the final state with respect to the initial state via a fluorescence detection\,\cite{Garttner2017BSN,Monroe2021CD}, similar as the many-body Loschmidt echo technique. For 2D lattice systems, the OTOC measurement is similar to that of the 1D lattice systems except for the construction of the model. Note that  our proposal is not limited to this particular architecture, and could be implemented or adapted in a variety of platforms that have full local quantum control\,\cite{Wang2018ZX,Cai2019HM,Li2017FW,Nie2020WC,Atala2013AB,Lohse2016SZ,Leseleuc2019LS,Wei2019PS,Zhao2022Ge}, such as a nuclear magnetic resonance quantum simulator\,\cite{Li2017FW,Nie2020WC,Wei2019PS} and superconducting qubit\,\cite{Wang2018ZX,Cai2019HM,Zhao2022Ge}.

In conclusion, we have proposed a zero-temperature OTOC witness in real space for identifying $\mathbb{Z}$-type TPTs in general lattice systems with or without chiral symmetry. Our proposal is robust to the choices of the initial state of the system and the used operators in OTOC.  It is also suitable for {\it disordered systems}, and can predict  topological Anderson insulator physics in the strong disorder regime. Moreover, the proposed OTOC witness can be used to detect not only first-order TPTs, but also second-order TPTs. Applying it into non-Hermitian systems\,\cite{supp}, the TPTs can be identified without implementing the transition from non-Bloch to Bloch theory. The generality of our proposal leads to that the proposed OTOC witness has predictive power in detecting TPTs. For example, we could construct the OTOC witness by preparing the system initially being in the first site and choosing a single-site operation as the $W$ operator, even in a situation where we don’t already understand the structure of a 1D lattice.

We thank Prof. T. Liu and Prof. J.-H. Gao for helpful discussions. This work is supported by the National Key Research and Development Program of China grant 2021YFA1400700, the National Science Foundation of China (Grants No.\,11974125, No.\,12205109, No.\,12147143), and the China Postdoctoral Science Foundation No. 2021M701323. F.N. is supported in part by:
Nippon Telegraph and Telephone Corporation (NTT) Research, the Japan Science and Technology Agency (JST) [via the Quantum Leap Flagship Program (Q-LEAP), and the Moonshot R\&D Grant Number JPMJMS2061], the Japan Society for the Promotion of Science (JSPS) [via the Grants-in-Aid for Scientific Research (KAKENHI) Grant No. JP20H00134], the Army Research Office (ARO) (Grant No. W911NF-18-1-0358), the Asian Office of Aerospace Research and Development (AOARD) (via Grant No. FA2386-20-1-4069), and the Foundational Questions Institute Fund (FQXi) via Grant No. FQXi-IAF19-06.

\onecolumngrid
\clearpage
\setcounter{equation}{0}
\setcounter{figure}{0}
\setcounter{table}{0}
\setcounter{page}{6}
\setcounter{section}{0}
\makeatletter
\renewcommand{\theequation}{S\arabic{equation}}
\renewcommand{\thefigure}{S\arabic{figure}}
\renewcommand{\bibnumfmt}[1]{[S#1]}
\renewcommand{\citenumfont}[1]{S#1}
\begin{center}
        \textbf{Supplemental Material for ``Out-of-Time-Order Correlation as a Witness for Topological Phase Transitions"}
\end{center}

\title{Supplemental Material for ``Nonreciprocal photon-phonon and photon-magnon pairs emission"}
\date{\today}

\title{Supplemental Material for ``Out-of-Time-Order Correlation as a Witness for Topological Phase Transitions"}
\author{Qian Bin}
\affiliation{School of Physics and Institute for Quantum Science and Engineering, Huazhong University of Science and Technology, Wuhan, 430074, China}

\author{Liang-Liang Wan}
\affiliation{School of Physics and Institute for Quantum Science and Engineering, Huazhong University of Science and Technology, Wuhan, 430074, China}

\author{Franco  Nori}
\affiliation{Theoretical Quantum Physics Laboratory, RIKEN Cluster for Pioneering Research, Wako-shi, Saitama 351-0198, Japan}
\affiliation{RIKEN Center for Quantum Computing (RQC), 2-1 Hirosawa, Wako-shi, Saitama 351-0198, Japan }
\affiliation{Physics Department, The University of Michigan, Ann Arbor, Michigan 48109-1040, USA}

\author{Ying Wu}
\affiliation{School of Physics and Institute for Quantum Science and Engineering, Huazhong University of Science and Technology, Wuhan, 430074, China}

\author{Xin-You L\"{u}}
\email{xinyoulu@hust.edu.cn}
\affiliation{School of Physics and Institute for Quantum Science and Engineering, Huazhong University of Science and Technology, Wuhan, 430074, China}
\date{\today}
\maketitle
This supplemental material contains five parts:  I. A detailed derivation of the analytical OTOC dynamics for the nearest-neighbor SSH model. II. Discussion of OTOC witness when the initial state is the eigenstate of the system. III. Additional discussion of the application of OTOC witness in the disordered systems. IV. Discussion of the application of OTOC witness in two-dimensional lattice described by the Qi-Wu-Zhang model. V. Discussion of the application of OTOC witness in the non-Hermitian systems.

\section{Derivation of the analytical OTOC dynamics}
In this section, we present the analytical derivation of the OTOC dynamics in the system described by the 1D  nearest-neighbor  Su-Schrieffer-Heeger (SSH)  model. We remind the Hamiltonian of the nearest-neighbor SSH model in the absence of disorder
\begin{align}\label{eq001}
H_{\rm s} = \sum_{n}\left\{ \epsilon \nu a_n^\dag \sigma_1 a_n +\frac{\epsilon}{2}\left[ a_{n+1}^\dag(\sigma_1+i\sigma_2)a_{n} + {\rm h.c.}   \right]   \right\},
 \end{align}
where $\sigma_j$ ($j=0,1,2,3$) are the Pauli matrices, corresponding to the identity matrix $I$, $\sigma_x$, $\sigma_y$, and $\sigma_z$, respectively. Here $a_n^\dag=(a_{n,A}^\dag, a_{n,B}^\dag)$ is the annihilation operator of the unit cell $n$ with sublattices $A$, $B$, and $\epsilon$ ($\epsilon\nu$) is the intercell (intracell) hopping strength. The parameter regimes $\nu<1$ and $\nu>1$ correspond to the topological non-trivial  and trivial phases, respectively.  This model has a chiral symmetry defined by a chiral operator $\mathcal{C}_{\rm 1d}=\sum_n a_n^\dag\sigma_3 a_n$ satisfying $[H_{\rm s},\mathcal{C}_{\rm 1d}]_+=0$. By defining $V=V\rho_0=|\psi_0\rangle\langle \psi_0|$, the constructed OTOC becomes an experimentally feasible fidelity of the final state $\rho_f$ projected onto an initial state $\rho_0$, with 
 \begin{align}\label{eq002}
    \mathcal{O}(t)={\rm tr}[ \rho_0 e^{i H_{\rm s}t}W^\dag  e^{-i H_{\rm s} t} \rho_0 e^{i H_{\rm s} t} We^{-i H_{\rm s}t}].
\end{align}
In the following, we address in turn the analytical solutions of the OTOC dynamics  for different  choices of the operator $W$ and the initial state $|\psi_0\rangle$.

We consider a general initial state 
 \begin{align}\label{eq003}
 |\psi_0\rangle=\frac{1}{\sqrt{M}}\sum_{m=1}^M (-1)^{m-1}|m,A\rangle,
 \end{align}
where $M=1$ corresponds to $|\psi_0\rangle=|1,A\rangle$. The matrix of the SSH Hamiltonian $H_{\rm s}$ including $2N$ sites in single-particle space has an implicit formula for the eigenpairs when $\nu \neq 1$\cite{S_Shin1997}. Then it is very difficult to  analytically calculate the OTOC dynamics under this Hamiltonian.  Comparing the cases of including $2N$ sites and $2N+1$ sites, the numerical energy spectrum of the latter is hardly changed except for one zero energy level is added, when the value of $N$ is large enough. The OTOC dynamics is almost not influenced by this extending of the Hamiltonian size. Thus, to analytically calculate the OTOC dynamics, we extend the size of the SSH model to $2N+1$ sites, and the Hamiltonian $H_{\rm s}$ becomes
\begin{gather}\label{eq004}
H_{\rm s}=\epsilon\left(\begin{array}{cccccc}
0 & \nu &  &  &  &  \\
\nu & 0 & 1 &  &   & \\
 & 1 & 0 & \nu & & \\
 & & \ddots & \ddots & \ddots & \\
 & & &  \nu & 0 & 1 \\ 
  & & &  & 1 & 0 
\end{array}\right)_{(2N+1)\times (2N+1)}.
\quad
\end{gather}
The eigenvalues of  Eq.\,(\ref{eq004}) are 
\begin{align}\label{eq005}
\lambda_{0}=0, ~~~~~
\lambda_\pm^{(k)}=\pm\epsilon \sqrt{1+\nu^2+2\nu  \cos\left(\frac{k\pi}{N+1}\right)},~~~~~~~~(k=1,2,\dots N)
\end{align}
and the corresponding eigenstates are
\begin{align}\label{eq006}
V^{(0)}=\frac{1}{\sqrt{A_0}}\left[(-\nu)^0, 0, (-\nu)^1, 0, (-\nu)^2, 0, \dots, 0, (-\nu)^{N-1}, 0, (-\nu)^N\right]^T
\end{align}
and 
\begin{align}\label{eq007}
V^{(k)}_{\pm}
=& \frac{1}{\sqrt{A_{\pm}^{(k)}}}\left[\frac{1}{\nu}\sin\left( \frac{0 k\pi}{N+1} \right)+\sin\left( \frac{ k\pi}{N+1} \right), \frac{\lambda^{(k)}_{\pm}}{\epsilon\nu}\sin\left(\frac{k\pi}{N+1}\right), \frac{1}{\nu}\sin\left( \frac{ k\pi}{N+1} \right)+\sin\left( \frac{ 2k\pi}{N+1} \right),   \right.\nonumber\\
&\left. \frac{\lambda^{(k)}_{\pm}}{\epsilon\nu}\sin\left(\frac{2k\pi}{N+1}\right), \frac{1}{\nu}\sin\left( \frac{2 k\pi}{N+1} \right)+\sin\left( \frac{ 3k\pi}{N+1} \right), \frac{\lambda^{(k)}_{\pm}}{\epsilon\nu}\sin\left(\frac{3k\pi}{N+1}\right),\dots,    \right.\nonumber\\
&\left. \frac{1}{\nu}\sin\left( \frac{(N-1) k\pi}{N+1} \right)+\sin\left( \frac{ Nk\pi}{N+1} \right), \frac{\lambda^{(k)}_{\pm}}{\epsilon\nu}\sin\left(\frac{Nk\pi}{N+1}\right), \frac{1}{\nu}\sin\left( \frac{N k\pi}{N+1} \right)+\sin\left( \frac{ (N+1)k\pi}{N+1} \right)
 \right]^T,
\end{align}
respectively, where
\begin{align}\label{eq008}
&A_0=\sum_{n'=0}^{N}\nu^{2n'}=\begin{cases}
(1-\nu^{2N+2})/(1-\nu^2) ~~~~~~~~~~~~(\nu\neq 1)\\
N+1 ~~~~~~~~~~~~~~~~~~~~~~~~~~~~~~~~(\nu=1)
\end{cases},~~~~~~~~\\
&A_{\pm}^{(k)}=\frac{(N+1)(\lambda^{(k)}_{\pm})^2}{\epsilon^2\nu^2}.
\end{align}
In the space spanned by the eigenstates $\{V^{(0)}, V^{(k)}_{\pm}\}$, the Hamiltonian $H_{\rm s}$ is given by 
 \begin{align}\label{eq010}
H_{\rm s}=\lambda^{(0)}|V^{(0)}\rangle \langle V^{(0)}|+\sum_{k=1}^N \left(\lambda_+^{(k)} |V_+^{(k)}\rangle\langle V_+^{(k)}|+ \lambda_-^{(k)} |V_-^{(k)}\rangle\langle V_-^{(k)}|\right).
 \end{align}
Then we obtain the  instantaneous state dominated by the forward evolution of $H_{\rm s}$, with
 \begin{align}\label{eq011}
&e^{-i H_{\rm s} t} |\psi_0\rangle = \exp\{-i [\lambda^{(0)}|V^{(0)}\rangle \langle V^{(0)}|+{\sum}_{k=1}^N (\lambda_+^{(k)} |V_+^{(k)}\rangle\langle V_+^{(k)}|+\lambda_-^{(k)} |V_-^{(k)}\rangle\langle V_-^{(k)}|)] t\}  |\psi_0\rangle\nonumber\\
 & = \frac{1}{\sqrt{MA_0}}\sum_{m=1}^M \nu^{m-1}   |V^{(0)}\rangle + \sum_{k=1}^N \sum_{m=1}^M\frac{(-1)^{m-1}}{\sqrt{M}}\left[\frac{1}{\nu}\sin(\frac{m-1}{N+1}k\pi)+\sin(\frac{mk\pi}{N+1})\right] \left[\frac{e^{-i\lambda_+^{(k)}t}}{\sqrt{A_+^{(k)}}}|V_+^{(k)}\rangle+\frac{e^{-i\lambda_-^{(k)}t}}{\sqrt{A_-^{(k)}}}|V_-^{(k)}\rangle\right].
 \end{align}
  where 
 \begin{align}\label{eq012}
&\langle V^{(0)}|m,A\rangle = \frac{1}{\sqrt{A_0}}(-\nu)^{m-1}, ~~~\langle V_{\pm}^{(k)}|m,A\rangle=\frac{1}{\sqrt{A_{\pm}^{(k)}}} \left[\frac{1}{\nu}\sin\left(\frac{m-1}{N+1}k\pi\right) + \sin\left(\frac{mk\pi}{N+1}\right)\right],
 \end{align}
and
 \begin{align}\label{eq013}
&\langle V^{(0)}|\psi_0\rangle = \frac{1}{\sqrt{MA_0}}(\nu^0+\nu^1+\nu^2+\dots+\nu^{M-1})= \frac{1}{\sqrt{MA_0}}\sum_{m=1}^M \nu^{m-1}, \\
&\langle V_{\pm}^{(k)}|\psi_0\rangle=\frac{1}{\sqrt{M A_{\pm}^{(k)}}} \sum_{m=1}^M (-1)^{m-1} \left[\frac{1}{\nu}\sin\left(\frac{m-1}{N+1}k\pi\right) + \sin\left(\frac{mk\pi}{N+1}\right)\right].
 \end{align}

 \begin{figure}
\includegraphics[width=17cm]{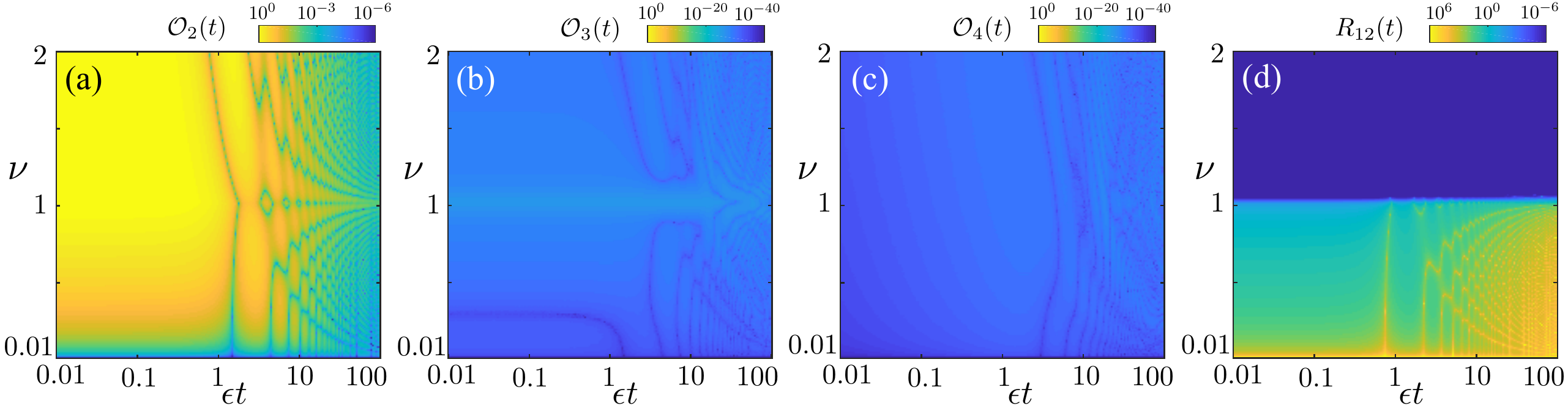}\\
\caption{ The values of (a) $\mathcal{O}_2(t)$,  (b) $\mathcal{O}_3(t)$,  (c) $\mathcal{O}_4(t)$, and (d) $R_{12}(t)$ versus $\epsilon t$ and $\nu$ when $W=\sum_{n=1}^{N-1} a_n^\dag \sigma_3 a_n$ and $|\psi_0\rangle=|1,A\rangle$. Other system parameters are $N=200$ and $d_1=d_2=0$.
}\label{sfig1}
\end{figure}

When  the OTOC operator $W=\sum_{l=1}^L a_l^\dag (\sigma_3+\sigma_0)a_l/2=\sum_{l=1}^La_{l,A}^\dag a_{l,A}$ ($L=1$ corresponds to the single-site operation), we obtain 
 \begin{align}\label{eq015}
 &\mathcal{O}(t)={\rm tr}[ \rho_0 e^{i H_{\rm s}t}W^\dag  e^{-i H_{\rm s} t} \rho_0 e^{i H_{\rm s} t} We^{-i H_{\rm s}t}]\nonumber\\
&\!\!=\left| \sum_{l=1}^L \!\left\{\! \frac{(-\nu)^{l-1}}{A_0\sqrt{M}}\!\!\sum_{m=1}^M \!\nu^{m-1}\!\!+\!\!\!\sum_{k=1}^N\!\sum_{m=1}^{M}\! \!\frac{2(-1)^{m-1}\!\cos(\lambda_+^{(k)}\!t)}{\sqrt{M}A_+^{(k)}}\! \left[\!\frac{1}{\nu}\sin(\frac{m\!-\!1}{N\!+\!1}k\pi)\!+\!\sin(\frac{mk\pi}{N\!+\!1})\!\right] \!\!\left[\!\frac{1}{\nu}\sin(\frac{l\!-\!1}{N\!+\!1}k\pi)\!+\!\sin(\frac{lk\pi}{N\!+\!1})\!\right] \! \right\}^2\right|^2.
 \end{align} 
This equation is reduced to Eq.\,(3) of the main text under the conditions of $L=1$ and $M=1$. When  the OTOC operator $W=\sum_{n=1}^{N-1} a_n^\dag \sigma_3 a_n=\sum_{n=1}^N a_n^\dag \sigma_3 a_n-a_N^\dag \sigma_3 a_N$, the OTOC function is reduced to
 \begin{align}\label{eq016}
\mathcal{O}(t)={\rm tr}[ \rho_0 e^{i H_{\rm s}t}W^\dag  e^{-i H_{\rm s} t} \rho_0 e^{i H_{\rm s} t} We^{-i H_{\rm s}t}]=|\mathcal{O}_1+\mathcal{O}_{2}(t)-\mathcal{O}_3(t)+\mathcal{O}_4(t)|^2
 \end{align}
 where
  \begin{align}\label{eq017}
&\mathcal{O}_1= \frac{1}{MA_0} \left( \sum_{m=1}^M \nu^{m-1}\right)^2,  \nonumber\\
&\mathcal{O}_2(t)=\sum_{k=1}^N\frac{2 \cos(2 \lambda_+^{(k)}t)}{MA_{\pm}^{(k)}}\left[\sum_{m=1}^M(-1)^{m-1}\left(\frac{1}{\nu}\sin(\frac{m-1}{N+1}k\pi)+\sin(\frac{mk\pi}{N+1})\right)\right]^2,\nonumber\\
&\mathcal{O}_3(t)=\left|\!\frac{(-\nu)^{N-1}}{A_0\sqrt{M}}\!\!\sum_{m=1}^M\!\nu^{m-1}\!+\!\!\sum_{k=1}^N\!\sum_{m=1}^M\!\frac{2(-1)^{m-1}\cos(2\lambda_{+}^{(k)}t)}{\sqrt{M}A_{\pm}^{(k)}}\![\frac{\sin(\frac{m-1}{N+1}k\pi)}{\nu}\!+\!\sin(\frac{mk\pi}{N+1})]\![\frac{\sin(\frac{N-1}{N+1}k\pi)}{\nu}\!+\!\sin(\frac{Nk\pi}{N+1})]\right|^2,\nonumber\\
&\mathcal{O}_4(t)=\left|\sum_{k=1}^N\!\sum_{m=1}^M\!\frac{-2i(-1)^{m-1}\lambda^{(k)}_+\sin(2\lambda_{+}^{(k)}t)}{\sqrt{M}A_{\pm}^{(k)}\epsilon \nu}\![\frac{1}{\nu}\!\sin(\frac{m-1}{N+1}k\pi)+\!\sin(\frac{mk\pi}{N+1})]\!\sin(\frac{Nk\pi}{N+1})\right|^2.\nonumber\\
 \end{align}
The terms $\mathcal{O}_2(t)$, $\mathcal{O}_3(t)$ and $\mathcal{O}_4(t)$ trend to zero over time, and $\mathcal{O}_3(t),\mathcal{O}_4(t)\ll\mathcal{O}_2(t)$ [see Figs.\,\ref{sfig1}(a-c)]. Then the above OTOC function can be approximately reduced to $\mathcal{O}(t)\approx|\mathcal{O}_1+\mathcal{O}_{2}(t)|^2$, and we can obtain the Eq.\,(4) of the main text under the conditions of $L=1$ and $M=1$. We define the ratio $R_{12}(t)=\mathcal{O}_1/\mathcal{O}_{2}(t)$. As shown in Fig.\,\ref{sfig1}(d), in the trivial phase $\nu>1$, $R_{12}(t)\ll 1$ means that the analytical solution can be approximatively reduced to $\mathcal{O}(t)\approx\mathcal{O}_2^2(t)$, which evolves to almost zero in the long-time limit. In the non-trivial phase $\nu<1$, we have $R_{12}(t\to \infty)\gg 1$ and the OTOC is reduced to $\mathcal{O}(t\to \infty)\approx\mathcal{O}^{2}_1$, which is a finite value. At the critical point $\nu=1$, $R_{12}(t\to \infty)\approx1$ corresponds to $\mathcal{O}(t\to \infty)\approx0$. This can also be seen in the phase diagram Fig.1(e) of the main text. Note that $\nu\neq 0$ in the above equations, and $\nu=0$ means that the hopping cannot occurs in the intercells. 
Here we have extended the size of the SSH model to $2N+1$ sites, but these solutions  Eqs.\,(\ref{eq015}-\ref{eq017})  are still valid for the system including $2N$ sites when the value of $N$ is large enough. In the main text, we have shown the numerical results obtained by numerically calculating Eq.\,(\ref{eq002}) for the system including $2N$ sites and the analytical results obtained by Eqs.\,(\ref{eq015}-\ref{eq017}) for the system including $2N+1$ sites, respectively. The excellent agreement between the analytical solutions and fully numerical simulations demonstrates the validity of our approximation. 

\begin{figure}
\includegraphics[width=16cm]{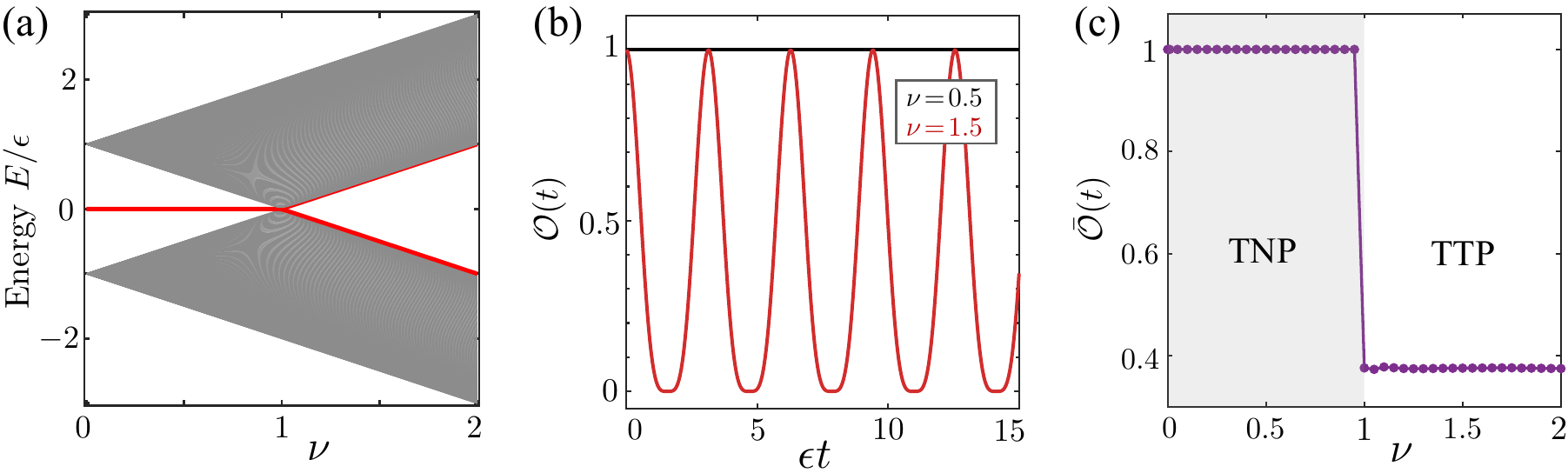}\\
\caption{(a) Energy spectrum of the 1D nearest-neighbor SSH model. (b) The evolution of the OTOC for different values of $\nu$, when the system is initially in the eigenstate $|\psi_E\rangle_{\rm 1d}$ of the system whose eigenvalue has the lowest absolute value, corresponding to the red curve of (a). (c) The average of the OTOC evolution  $\bar{\mathcal{O}}(t)$  versus $\nu$.  System parameters are $N=200$ and $W=\sum_{n=1}^N a_{n,A}^\dag a_{n,A}$. The gray and white areas correspond to the topological non-trivial phase (TNP) and topological trivial phase (TTP), respectively.}\label{sfig2}
\end{figure}
\begin{figure}
\includegraphics[width=12cm]{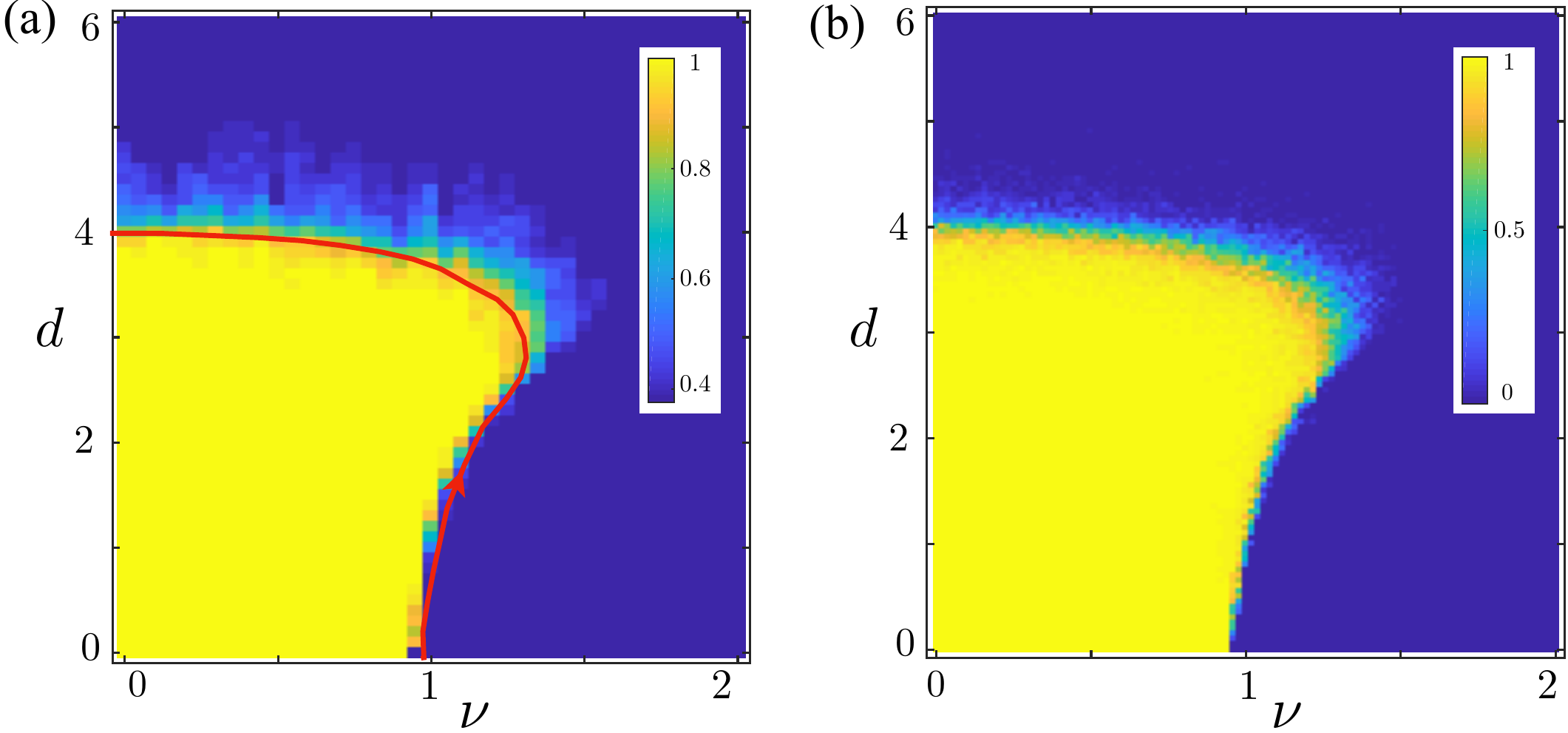}\\
\caption{Topological phase diagrams: (a) the averaged OTOC $\bar{\mathcal{O}}(t)$ versus disorder strength $d$ and $\nu$ when $W=\sum_{n=1}^N a_{n,A}^\dag a_{n,A}$ and  $|\tilde{\psi}_{0}\rangle=Q_{\rm 1d}|\psi_{E}\rangle_{\rm 1d}$; (b) the order parameter defined in Refs.\,\cite{S_MondragonShem2014HS,S_Meier2018AD} versus disorder strength $d$ and $\nu$. Here all data are averaged over 30 independent disorder configurations, and $d_2=2d_1=d$ and $N=1000$.
}\label{sfig3}
\end{figure}

\section{The OTOC witness when the initial state is the eigenstate of system}
In the section, we discuss the connection between OTOC dynamics and TPTs when the initial state is the eigenstate $|\psi_E\rangle_{\rm 1d}$ of the system whose eigenvalue has the lowest absolute value [see the red curve in Fig.\,\ref{sfig2}(a)]. Let's take the system described by the the nearest-neighbor SSH model as an example. Applying an operation on the initial state, we obtain $|\tilde{\psi}_{0}\rangle=Q_{\rm s}|\psi_{E}\rangle_{\rm 1d}$, where $Q_{\rm s}=\sum_{n}a_n^\dag(\sigma_0+\sigma_3)a_n/2$. In the topological non-trivial phase $\nu<1$, the state $|\psi_{E}\rangle_{\rm 1d}$ is naturally an edge state of the system, e.g., the left edge state $\sum_{n}\alpha_{n}|n,A\rangle$ with $\sum_n|\alpha_n|^2=1$, and then $|\tilde{\psi}_{0}\rangle=Q_{\rm 1d}|\psi_{E}\rangle_{\rm 1d}=\sum_{n}\alpha_{n}|n,A\rangle=|\psi_{E}\rangle_{\rm 1d}$. The OTOC is reduced to $\mathcal{O}(t)  ={\rm tr}[\tilde{\rho}_{0}^{2}]=1$, which is a conserved quantity. The averaged OTOC function $\bar{\mathcal{O}}(t)=1$. In the topological trivial phase $\nu>1$, $|\psi_{E}\rangle_{\rm 1d}$ is not the edge state of the system, but a bulk state $|\psi_{E}\rangle_{\rm 1d}=\sum_{n}(\alpha_{n}|n,A\rangle +\beta_{n} |n,B\rangle)$, where $\sum_n (|\alpha_{n}|^2+|\beta_n|^2)=1$ and $\sum_n |\alpha_{n}|^2=\sum_n |\beta_{n}|^2=1/2$. In Figs.\,\ref{sfig2}(b,c), we show the OTOC dynamics and averaged OTOC in different topological phases when the OTOC operator $W=\sum_{n=1}^N a_{n,A}^\dag a_{n,A}=\sum_{n=1}^Na_{n}^\dag(\sigma_0+\sigma_3)a_n/2$ and  $|\tilde{\psi}_{0}\rangle=Q_{\rm s}|\psi_{E}\rangle_{\rm 1d}$. The results are obtained by numerically calculating the OTOC function Eq.\,(\ref{eq002}), which demonstrates that the  OTOC periodically oscillates in the topological trivial phase and becomes a conserved quantity in the topological non-trivial phase. The averaged OTOC $\bar{\mathcal{O}}(t)$ is thus discrete at the critical point, i.e.,  $\bar{\mathcal{O}}(t)=0.375$ and  $\bar{\mathcal{O}}(t)=1$ in the trivial and non-trivial phases, respectively.

\section{Additional discussion of the application of OTOC witness in the disordered systems}
We remind the Hamiltonian of the disordered nearest-neighbor  SSH model 
\begin{align}\label{eq022}
H_{\rm s} = \sum_{n}\left\{ \nu_n a_n^\dag \sigma_1 a_n +\frac{\omega_n}{2}\left[ a_{n+1}^\dag(\sigma_1+i\sigma_2)a_{n} + {\rm h.c.}   \right]   \right\},
 \end{align}
where $\omega_n=\epsilon(1+d _1r_n)$ [or $\nu_n=\epsilon(\nu+d_2 r_n')$] is the intercell (or intracell) hopping strength. Disorder with the dimensionless strengths $d_1$, $d_2$ has been included here, and $r_n$, $r_n'$ are independent random real numbers chosen from the uniform distribution $[-0.5,0.5]$. In the clean system (i.e., $d_1=d_2=0$), the above equation is reduced to a standard SSH Hamiltonian Eq.\,(\ref{eq001}).  Indeed, the symmetry-protected boundary state has strong robustness to weak disorder, but the topological features disappear as the disorder is too large.  Moreover, the disorder can also induce the appearance of the non-trivial topology when it is added in a topological trivial  structure. This disorder-driven topological phase is called as topological Anderson insulator phase.  To fully show the effects of disorder on the TPTs, we numerically calculate the OTOC function Eq.\,(\ref{eq002}) with different system parameters by choosing the OTOC operator $W=\sum_{n=1}^N a_{n,A}^\dag a_{n,A}$ and  $|\tilde{\psi}_{0}\rangle=Q_{\rm 1d}|\psi_{E}\rangle_{\rm 1d}$. Figure\,\ref{sfig3}(a) displays the dependence of the averaged OTOC $\bar{\mathcal{O}}(t)$ on the disorder strength $d$ and $\nu$, which can be considered as a topological phase diagram for the systems including disorders. There still exists an obvious step transition from $0.375$ to $1$ at the critical point in the presence of weak disorder. When the disorder is increased, the distinguishability of the OTOC dynamics disappear. Besides confirming the robustness of the OTOC witness to weak disorder, this phase diagram also shows topological Anderson insulator phase. For example, the system enters into the topological non-trivial phase from the trivial phase along with increasing the disorder strength, when the value of $\nu$ is slightly larger than 1. The physical mechanism for this result can be explained as follows. The relative strong disorder can induce an addition locality on the system, which leads a shift of the critical point of TPTs, as the red curve in Fig.\,\ref{sfig3}(a). Then, in the phase diagram, there appears a range of $\nu$ corresponding to the occurrence of the TPT from the trivial to non-trivial phase with increasing disorder. Figure\,\ref{sfig3}(b) shows the similar phase diagram from Refs.\,\cite{S_MondragonShem2014HS,S_Meier2018AD}. Here, the topological phase diagram obtained by the OTOC witness is consistent with previous works, which verify the validity of our results. 

\section{The application of OTOC witness in two-dimensional lattice described by the Qi-Wu-Zhang model}
\begin{figure}
\includegraphics[width=13cm]{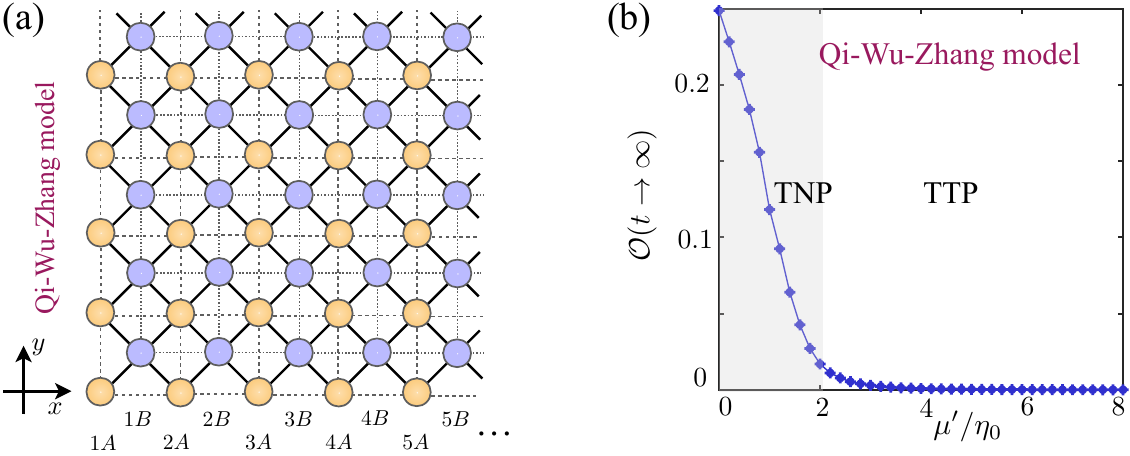}\\
\caption{ (a) Scheme of the Qi-Wu-Zhang model. The unit cell consists of sublattices $A$ and $B$. 
(b) The OTOC in the long-time limit $\mathcal{O}(t\to\infty)$ versus  $\mu'/\eta_0$. The TNP and TTP are indicated by the gray shading and write area, respectively. The cell numbers of $x$ and $y$ directions $N_x=N_y=20$,  and other system parameters are $W=\sum_{x,y} a_{x,y}^\dag (\sigma_0-\sigma_3) a_{x,y}^\dag/2$ and $|\psi_0\rangle=|1,A\rangle$.}\label{sfig4}
\end{figure}

In this section, we discuss the connection between OTOC dynamics and TPTs of 2D lattice system described by the Qi-Wu-Zhang model.  As shown in Fig.\,\ref{sfig4}(a), the Qi-Wu-Zhang model on the 2D square lattice has Hamiltonian\,\cite{S_Qi2006WZ,S_Asboth2016OP}
 \begin{align}\label{eq023}
H_{\rm qwz}=\sum_{x,y} \left\{\eta_0\left[ a_{x+1,y}^\dag \left( \frac{\sigma_3+i\sigma_1}{2}\right) a_{x,y}+a_{x,y+1}^\dag \left( \frac{\sigma_3+i\sigma_2}{2}\right) a_{x,y} +{\rm h.c.} \right]+\mu' a_{x,y}^\dag \sigma_3 \sigma_{x,y}\right\},
\end{align}
where $a_{x,y}^\dag=(a_{x,y,A}^\dag, a_{x,y,B}^\dag)$ is the creation operator of the unit cell with sublattices $A$ and $B$, $\eta_0$ is the hopping strength, and $\mu'$ is the sublattice potential.  The system does not have chiral symmetry. The topological  trivial and non-trivial phases in the system can be identified by Chern number in momentum space. The parameter ranges $-2<\mu'/\eta_0<2$ and $\mu'/\eta_0= other$ correspond to topological trivial and non-trivial phases, respectively. In Fig.\,\ref{sfig4}(b), we show that the OTOC of the system in the long-time limit $\mathcal{O}(t\to\infty)$ as a function of $\mu'/\eta_0$ when $W=\sum_{x,y} a_{x,y}^\dag (\sigma_0- \sigma_3) a_{x,y}/2$ and $|\psi_0\rangle=|1,A\rangle$, where $[W,H_{\rm qwz}]_{\pm}\neq0$. The distinct behavior of the OTOC dynamics in the trivial and non-trivial phases is still observed in the system. The value of $\mathcal{O}(t\to\infty)$ changes from almost zero to the finite values, when the system enters into the non-trivial phase from the trivial phase.

\section{The application of OTOC witness in non-Hermitian systems}
In this section, we discuss the connection between TPTs and OTOC dynamics in non-Hermitian systems\,\cite{S_Leykam2017BH,S_Gneiting2020KV,S_Minganti2021AM,S_Leefmans2022DW,S_Jin2021Song}. The Hamiltonian of the 1D non-Hermitian SSH model reads\,\cite{S_Lieu2018,S_Chen2019DS,S_Kunst2018EB,S_Yao2018Wang}
\begin{align}\label{eq024}
H_{\rm nh} = \sum_{n}\left\{\frac{\epsilon (\nu+\delta)}{2} a_n^\dag (\sigma_1+i\sigma_2) a_n +\frac{\epsilon (\nu-\delta)}{2} a_n^\dag (\sigma_1-i\sigma_2) a_n +\frac{\epsilon}{2}\left[ a_{n+1}^\dag(\sigma_1+i\sigma_2)a_{n} + {\rm h.c.}   \right]   \right\},
\end{align}
where $\sigma_j$ ($j=0,1,2,3$) are the Pauli matrices, corresponding to the identity matrix $I$, $\sigma_x$, $\sigma_y$, and $\sigma_z$, respectively. Here $a_n^\dag=(a_{n,A}^\dag, a_{n,B}^\dag)$ is the annihilation operator of the unit cell $n$ with sublattices $A$, $B$, and $\epsilon$ is the intercell hopping strength,  and $\epsilon(\nu+\delta)$ and  $\epsilon(\nu-\delta)$ are the intracell hopping strength. When $\delta=0$, Eq.\,(\ref{eq024}) is reduced to the standard SSH Hamiltonian\,(\ref{eq001}), with the phase transition point $\nu_c=1$. When $\delta\neq0$,  Eq.\,(\ref{eq024}) is a non-Hermitian Hamiltonian, and the phase transition point is $\nu_c=\sqrt{1+\delta^2}$, where $\nu<\nu_c$ and $\nu>\nu_c$ correspond to the topological non-trivial and trivial phases, respectively. 

\begin{figure}
\includegraphics[width=13cm]{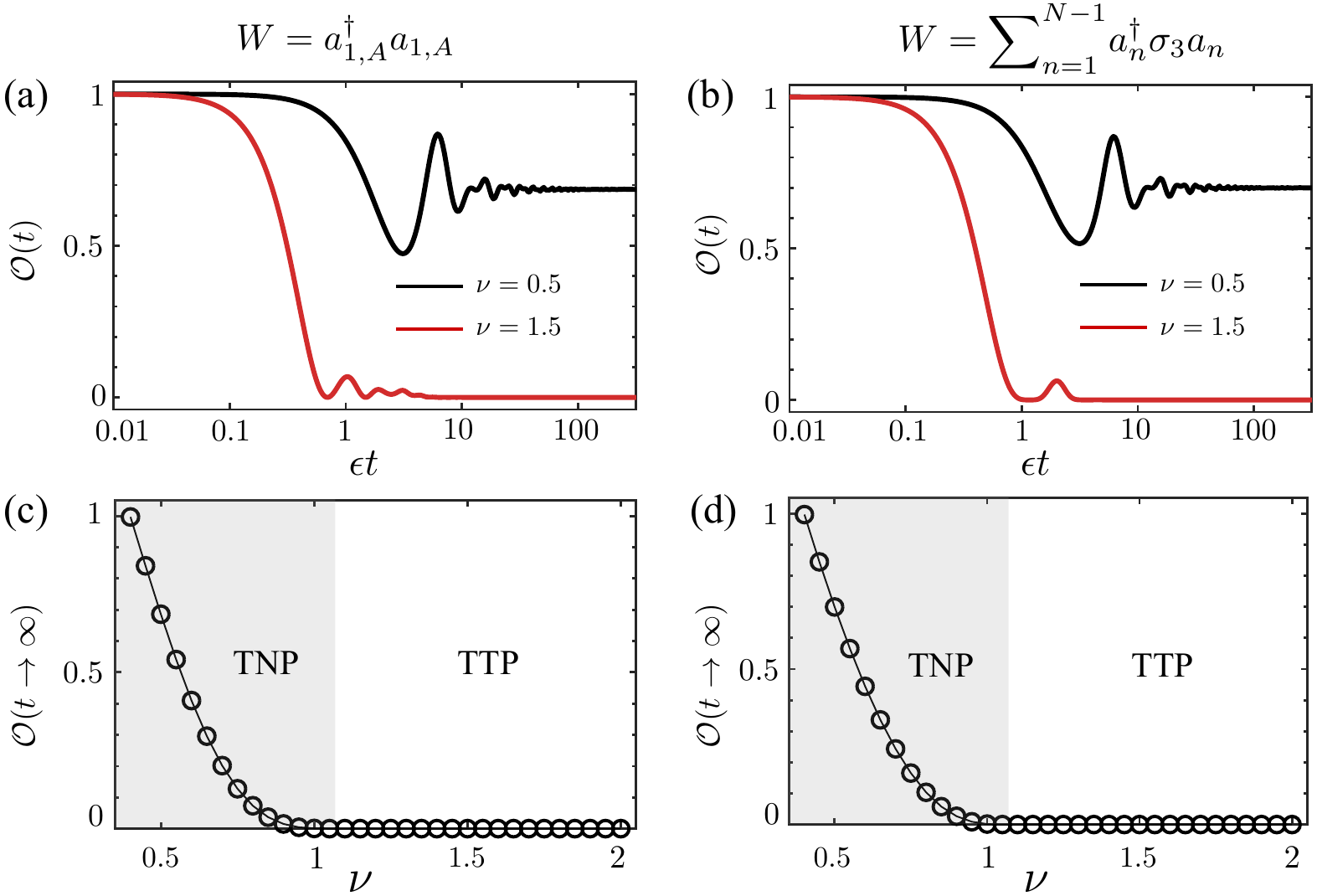}\\
\caption{(a--b) Dominated by $H_{\rm nh}$, the evolution of the OTOC for different values of $\nu$. (c--d) The OTOC in the long-time limit $\mathcal{O}(t\rightarrow\infty)$ versus $\nu$. The gray and white areas correspond to the TNP and TTP, respectively.  Other system parameters are $\delta=0.4$, $|\psi_0\rangle=|1,A\rangle$, (a,c) $W=a_{1,A}^\dag a_{1,A}$, and (b,d) $W=\sum_{n=1}^{N-1} a_{n}^\dag \sigma_3 a_{n}$.
}\label{sfig5}
\end{figure}

Similar to the discussion in Hermitian systems, here we choose  $|\psi_0\rangle=|1,A\rangle$, and take the OTOC operators $W=a_{1,A}^\dag a_{1,A}$ and  $W=\sum_{n=1}^{N-1} a_{n}^\dag \sigma_3 a_{n}$ as examples, where $|n,A/B\rangle$  represents the system occupying in the sublattice $A/B$ of the unit cell $n$. As shown in Figs.\,\ref{sfig5}(a,b), the OTOC becomes almost zero along with the time evolution in the trivial phase, while it trends to a non-zero finite value in the topological non-trivial phase. This means that the distinguished OTOC dynamics in the trivial and non-trivial phases can be obtained for two choices of the operator $W$. In Figs\,\ref{sfig5}(c,d), we present the OTOC in the long-time limit $\mathcal{O}(t\rightarrow\infty)$ during a wide range of parameters, which shows a sudden change of the values of $\mathcal{O}(t\rightarrow\infty)$ at the phase transition point. The above results further demonstrate that the constructed OTOC can still be a witness for detecting TPTs even in non-Hermitian systems.

\end{document}